\documentstyle[12pt]{article}

\catcode`\@=11
\long\def\@makefntext#1{
\protect\noindent \hbox to 3.2pt {\hskip-.9pt
$^{{\ninerm\@thefnmark}}$\hfil}#1\hfill}                

 \def\@makefnmark{\hbox to 0pt{$^{\@thefnmark}$\hss}}  

\def\ps@myheadings{\let\@mkboth\@gobbletwo
\def\@oddhead{\hbox{}
\rightmark\hfil\ninerm\thepage}
\def\@oddfoot{}\def\@evenhead{\ninerm\thepage\hfil
\leftmark\hbox{}}\def\@evenfoot{}
\def\sectionmark##1{}\def\subsectionmark##1{}}

\newcounter{sectionc}\newcounter{subsectionc}\newcounter{subsubsectionc}
\renewcommand{\section}[1] {\vspace{0.6cm}\addtocounter{sectionc}{1}
\setcounter{subsectionc}{0}\setcounter{subsubsectionc}{0}\noindent
        {\bf\thesectionc. #1}\par\vspace{0.4cm}}
\renewcommand{\subsection}[1] {\vspace{0.6cm}\addtocounter{subsectionc}{1}
        \setcounter{subsubsectionc}{0}\noindent
        {\it\thesectionc.\thesubsectionc. #1}\par\vspace{0.4cm}}
\renewcommand{\subsubsection}[1]
{\vspace{0.6cm}\addtocounter{subsubsectionc}{1}
        \noindent {\rm\thesectionc.\thesubsectionc.\thesubsubsectionc.
        #1}\par\vspace{0.4cm}}

\newcounter{appendixc}
\newcounter{subappendixc}[appendixc]
\newcounter{subsubappendixc}[subappendixc]

\renewcommand{\appendix}[1] {\vspace{0.6cm}
        \refstepcounter{appendixc}
        \setcounter{figure}{0}
        \setcounter{table}{0}
        \setcounter{equation}{0}
        \renewcommand{\thefigure}{\Alph{appendixc}.\arabic{figure}}
        \renewcommand{\thetable}{\Alph{appendixc}.\arabic{table}}
        \renewcommand{\theappendixc}{\Alph{appendixc}}
        \renewcommand{\theequation}{\Alph{appendixc}.\arabic{equation}}
        \noindent{\bf Appendix \theappendixc #1}\par\vspace{0.4cm}}

\def\abstracts#1{{
        \centering{\begin{minipage}{30pc}\tenrm\baselineskip=12pt\noindent
        \centerline{\tenrm ABSTRACT}\vspace{0.3cm}
        \parindent=0pt #1
        \end{minipage}}\par}}

\renewenvironment{thebibliography}[1]
        {\begin{list}{\arabic{enumi}.}
        {\usecounter{enumi}\setlength{\parsep}{0pt}
\setlength{\leftmargin 0.52cm}{\rightmargin 0pt}
         \setlength{\itemsep}{0pt} \settowidth
        {\labelwidth}{#1.}\sloppy}}{\end{list}}

\newcounter{itemlistc}
\newcounter{romanlistc}
\newcounter{alphlistc}
\newcounter{arabiclistc}

\newcommand{\fcaption}[1]{
        \refstepcounter{figure}
        \setbox\@tempboxa = \hbox{\tenrm Fig.~\thefigure. #1}
        \ifdim \wd\@tempboxa > 6in
           {\begin{center}
        \parbox{6in}{\tenrm\baselineskip=12pt Fig.~\thefigure. #1}
            \end{center}}
        \else
             {\begin{center}
             {\tenrm Fig.~\thefigure. #1}
              \end{center}}
        \fi}

\newcommand{\tcaption}[1]{
        \refstepcounter{table}
        \setbox\@tempboxa = \hbox{\tenrm Table~\thetable. #1}
        \ifdim \wd\@tempboxa > 6in
           {\begin{center}
        \parbox{6in}{\tenrm\baselineskip=12pt Table~\thetable. #1}
            \end{center}}
        \else
             {\begin{center}
             {\tenrm Table~\thetable. #1}
              \end{center}}
        \fi}

\def\@citex[#1]#2{\if@filesw\immediate\write\@auxout
        {\string\citation{#2}}\fi
\def\@citea{}\@cite{\@for\@citeb:=#2\do
        {\@citea\def\@citea{,}\@ifundefined
        {b@\@citeb}{{\bf ?}\@warning
        {Citation `\@citeb' on page \thepage \space undefined}}
        {\csname b@\@citeb\endcsname}}}{#1}}

\newif\if@cghi
\def\cite{\@cghitrue\@ifnextchar [{\@tempswatrue
        \@citex}{\@tempswafalse\@citex[]}}
\def\citelow{\@cghifalse\@ifnextchar [{\@tempswatrue
        \@citex}{\@tempswafalse\@citex[]}}
\def\@cite#1#2{{$\null^{#1}$\if@tempswa\typeout
        {IJCGA warning: optional citation argument
        ignored: `#2'} \fi}}

\def\fnt#1#2{\footnotetext{\kern-.3em
        {$^{\mbox{\sevenrm #1}}$}{#2}}}

 1
 1
 1

\font\tenbf=cmbx10
\font\tenrm=cmr10
\font\tenit=cmti10

\font\ninerm=cmr9

\input{epsf}

\begin{document}
\vspace*{6cm}
\begin{center}
  \begin{Large}
  \begin{bf}
THE BESS MODEL AT FUTURE COLLIDERS$^*$\\
  \end{bf}
  \end{Large}
  \vspace{1.5cm}
  \begin{large}
Aldo Deandrea\\
  \end{large}
\vspace{4mm}
D\'epartement de Physique Th\'eorique, Univ. de Gen\`eve\\
24, quai Ernest-Ansermet - CH-1211 Gen\`eve 4\\
\end{center}
\vspace{4cm}
\begin{center}
UGVA-DPT 1994/11-867\\\
hep-ph/9411291\\\
November 1994
\end{center}
\vspace{2cm}
\noindent
$^*$ Contribution to the proceedings of the First Arctic Workshop on
Future Physics and Accelerators, Saariselk\"a, Finland, August 1994.
Partially supported by the Swiss National Foundation.
\newpage

\def\lq{\left [}
\def\rq{\right ]}
\def\qq{Q^2}
\def\dmu{\partial_{\mu}}
\def\dmus{\partial^{\mu}}
\def\AA{{\cal A}}
\def\BB{{\cal B}}
\def\Tr{{\rm Tr}}
\def\gp{g'}
\def\gs{g''}
\def\ggs{\frac{g}{\gs}}
\def\mpp{M_{P^+}}
\def\mpm{M_{P^-}}
\def\mpt{M_{P^3}}
\def\mpz{M_{P^0}}
\def\eps{{\epsilon}}
\def \LL{{{\bf L}_\mu}}
\def \RR{{{\bf R}_\mu}}
\newcommand{\be}{\begin{equation}}
\newcommand{\ee}{\end{equation}}
\newcommand{\bea}{\begin{eqnarray}}
\newcommand{\eea}{\end{eqnarray}}
\newcommand{\nn}{\nonumber}
\newcommand{\dd}{\displaystyle}

\centerline{\tenbf THE BESS MODEL AT FUTURE COLLIDERS}
\baselineskip=16pt
\vspace{0.8cm}
\centerline{\tenrm Aldo Deandrea}
\baselineskip=13pt
\centerline{\tenit D\'epartement de Physique Th\'eorique, Universit\'e de
Gen\`eve}
\baselineskip=12pt
\centerline{\tenit 24, quai Ernest-Ansermet - CH-1211 Gen\`eve 4 - Suisse}
\vspace{0.9cm}
\abstracts{The BESS model consists of an effective lagrangian parametrization
with dynamical symmetry breaking, describing scalar, vector and axial-vector
bound states in a rather general framework.  After a brief description of the
model and its generalizations, predictions for physics at future accelerators
are given both for LHC and $e^+e^-$ machines.}

\vfil
\rm\baselineskip=14pt
\section{Introduction}

The mechanism for symmetry breaking in the standard model is usually regarded
as unsatisfactory. More fundamental realizations lead to
the expectation that higher energy accelerators might reveal new particles and
interactions. Our parametrization in terms of scalar couplings may in fact
represent the effective low energy manifestation of more
fundamental dynamics, with additional particles and interactions.

The fundamental dynamics may have the form of a new strong interaction
\cite{techni} \cite{extechni}, but the construction of a satisfactory
technicolor theory is a difficult task and in the absence of a specific
theory of the strong electroweak sector, one would like to remain as general
as possible, avoiding specific dynamical assumptions. The BESS model
\cite{BESS} (Breaking Electroweak Symmetry Strongly) was essentially
developed to provide for such a general frame (for a review see \cite{rev}).

\section{BESS}

In the standard model (SM) the symmetry breaking is
realized linearly with scalars originally transforming as the $({1 \over 2}, {1
\over 2})$ of $SU(2)_L\otimes SU(2)_R$. The direct product breaks into
$SU(2)_{diagonal} $, with corresponding breaking of $({1 \over 2}, {1 \over
2})$
into $1 \oplus 3$, describing the physical Higgs and the 3 absorbed Goldstones.

The non-linear realization of symmetry breaking I shall consider in the
following, can be seen classically to correspond to the limit of infinite
$m_H$. The scalars can be represented as proportional to a unitary
matrix $U$. In the formal limit $m_H \rightarrow \infty$ one is freezing
the proportionality factor to the vacuum expectation value and the scalar
lagrangian is the lagrangian of a non linear $\sigma$-model described by a
unitary matrix $U$, invariant under $SU(2)_L \otimes SU(2)_R$,
namely under $U\rightarrow g_L Ug_R^{\dagger}$ where $g_L, g_R$ belong to
$SU(2)_L, SU(2)_R$ respectively. The breaking into the diagonal $SU(2)$
is demanded by the non-linear unitarity condition $U^{\dagger}U=1$.

The idea is to construct a model with vector and axial-vector particles, so
that when these particles decouple one obtains the non linear $\sigma$-model
lagrangian. Indeed, one may consider a more general case and start from a
global symmetry $SU(N)_L\otimes SU(N)_R$ \cite{bessu8}, rather than
$SU(2)_L\otimes SU(2)_R$ \cite{BESS}. In order to do that, one introduces a
local copy of the global symmetry, $[SU(N)_L\otimes SU(N)_R]_{local}$.
When the new vector and axial-vector particles decouple, one
obtains the non-linear $\sigma$-model lagrangian, describing the
Goldstone bosons, transforming as the representation $(N,N)$
of $SU(N)_L\otimes SU(N)_R$, associated to the breaking of
$SU(N)_L\otimes SU(N)_R\to SU(N)_{L+R}$
\be
{\cal L} = {v^2\over 2N} Tr \left [(\partial_\mu U)(\partial^\mu U)^\dagger
\right ]
\label{sigma}
\ee
To introduce both vector and axial-vector particles, I assume the following
factorization of $U$
\be
U=LM^\dagger R^\dagger
\ee
where $L,~M,~R$, transform according to the representations of
\be
G=[SU(N)_L\otimes SU(N)_R]_{global}\otimes[SU(N)_L\otimes
SU(N)_R]_{local}\nn
\ee
as
\def\spin{{N}}
\be
L\in (\spin,0,\spin,0)~~~~~
M\in (0,0,\spin,\spin)~~~~~
R\in (0,\spin,0,\spin)
\ee
which means
\be
L^\prime = g_L L h_L~~~~~
M^\prime = h_R^\dagger M h_L~~~~~
R^\prime = g_R R h_R
\ee
where
\bea
& & g_L\in ({SU(N)_L})_{global}~~~~~
g_R\in ({SU(N)_R})_{global}\nn\\
 & &h_L\in ({SU(N)_L})_{local}~~~~~
h_R\in ({SU(N)_R})_{local}
\eea
In this way $U$ does not transform under the local symmetry
(hidden gauge symmetry \cite{bando}):
\be
U^\prime = g_L U g_R^\dagger
\ee
that is,
\be
U\in (\spin,\spin,0,0)
\ee
The Lagrangian in Eq.(\ref{sigma}) is invariant under the discrete
transformation $U\to U^\dagger$, which corresponds to
(parity transformation):
\be
L\to R~~~~~
M\to M^\dagger~~~~~
R\to L
\label{par}
\ee

Covariant derivatives can be built with respect to the local group:
\bea
& &D_\mu L =\partial_\mu L - L\LL \nn \\
& &D_\mu R =\partial_\mu R - R\RR \nn \\
& &D_\mu M =\partial_\mu M - M\LL + \RR M
\eea
where $\LL$ and $\RR$ are the Lie algebra valued gauge fields of
$({SU(N)_L})_{local}$ and $({SU(N)_R})_{local}$ respectively.

The invariants of our original group extended by the
parity operation defined in Eq.(\ref{par}) are:
\bea
{I}_1&=&\Tr ({L}^\dagger D_\mu {L}
       -M^\dagger D_\mu M-M^\dagger {R}^\dagger
          (D_\mu  {R}) M)^2\\
{I}_2&=&\Tr ({L}^\dagger D_\mu {L}+M^\dagger
           {R}^\dagger (D_\mu {R}) M)^2\\
{I}_3&=&\Tr ({L}^\dagger D_\mu {L}-M^\dagger {R}
         ^\dagger (D_\mu {R}) M)^2\\
{I}_4&=&\Tr (M^\dagger D_\mu M)^2
\eea
Using these invariants I can write the most general Lagrangian with at
most two derivatives in the form:
\be
{\cal L}=-\frac{v^2}{2N} (a {I}_1+b {I}_2+c {I}_3
+d {I}_4)+~kinetic~terms~for~the~gauge~fields
\label{lb}
\ee
where $a,~b,~c,~d$ are free parameters and furthermore the gauge coupling
constant for the fields $\LL$ and $\RR$ is the same.

The requirement of getting back to the non-linear $\sigma$-model in
the limit in which the gauge fields $\LL$ and $\RR$ are decoupled is
satisfied by imposing the relation $a+ z d =1$ where
\be
z=\frac{c}{c+d}
\ee
The case $z=0$ corresponds to the decoupling of the axial-vector resonances.

Gauging the previous effective Lagrangian with respect to the standard
gauge group $SU(3)_c\otimes SU(2)_L\otimes U(1)_R$:
\bea
D_\mu {L} &\to &  D_\mu {L} =
\dmu {L} -{L}
            (V_\mu-A_\mu)+{\bf A}_\mu {L}\\
D_\mu {R} &\to & D_\mu {R} = \dmu {R} -{R}
           (V_\mu+A_\mu)+{\bf B}_\mu {R}\\
D_\mu M &\to & D_\mu {M} =\dmu M -M (V_\mu-A_\mu)+(V_\mu+A_\mu) M
\eea
where $V_\mu=(\RR+\LL)/2$ and $A_\mu=(\RR-\LL)/2$ are the fields describing the
new vector and axial-vector resonances, whereas ${\bf A}_\mu$ and ${\bf B}_\mu$
are linear combinations of the gauge fields of the standard gauge group.

In the following I shall consider the two cases $N=2$ and
$N=8$, to which the quantitative discussion will be limited, except for
some occasional more general remark. The $SU(2)$ case has 3 new vector and 3
new axial-vector resonances. The $SU(8)$ case is obviously much richer and can
be specialized to standard $SU(8)$ technicolor; in this case also spin-0
pseudo-Goldstone particles are present in the spectrum.

I shall denote the $SU(8)$ gauge fields as $V^A=(V^a,{\tilde
V}^a,V_D,V_8^\alpha,V_8^{a \alpha},V_3^{\mu i}, {\bar V}_3^{\mu i})$, where
$\mu=(0,a)$ ($a$ being an $SU(2)$ index), $\alpha$ an octet index and
$i=1,2,3$ is a color index. An analogous notation will be used for $A^A$ and
the Goldstone bosons $\pi^A$. In the following I shall use the notations:
\begin{eqnarray}
{\bf A}_\mu &=&
2 i g W^a_\mu T^a+i \sqrt{2} g_s G^\alpha_\mu T^\alpha_8+2 i y\gp
Y_\mu T_D\\
{\bf B}_\mu &=&
2 i \gp Y_\mu(T^3 + y T_D)+
i  \sqrt{2} g_s G^\alpha_\mu T^\alpha_8\\
V_\mu &=& i \gs V_\mu^A T^A \\
A_\mu &=& i\gs A_\mu^A T^A
\end{eqnarray}
with $A,B=1,\ldots ,63$; $a,b=1,2,3$; $\alpha ,\beta =1,\ldots ,8$ and $y
=1/\sqrt{3}$; $W,~Y,~G$ are the standard model gauge bosons and $g,~\gp,g_s$
their coupling constants, while $\gs$ is the self coupling of the $V$ and $A$
bosons. The $V$ and $A$ bosons can be decoupled by sending $\gs\to\infty$.
In this limit, the mass of the $W$ bosons is the SM mass with $v\simeq
246~GeV$.

\section{$e^+e^-$ future colliders}

Future $e^+ e^-$ linear colliders with different centre of mass energies and
luminosities have been proposed; a collider with energy up to 500 GeV
has concentrated most of the studies \cite{ee500}, but at the same time
possibilities of centre of mass energies of 1 or 2 TeV have been discussed
\cite{saari}.

The BESS model, even in its minimal formulation, contains new vector
resonances. If the mass $M_V$ of the new boson
multiplet lies not far from the maximum machine energy, or if it is lower,
such a resonant contribution will be manifest.

One can measure the fermionic channel $e^+e^- \to f
{\bar f}$, but this will not give a real improvement with respect to the
existing bounds from LEP1. In Fig.~1 restrictions on the parameter space of the
model from LEP1 data are obtained in the low energy limit of the BESS model.
The method consists in eliminating the heavy vector field
by the use of its classical equations of motion, in the infinite mass limit.
After the elimination of the heavy degrees of freedom the additional
terms to the Standard Model lagrangian allow to read directly the deviations
from the Standard Model \cite{ste}.
\begin{figure}
\epsfysize=11truecm
\centerline{\epsffile{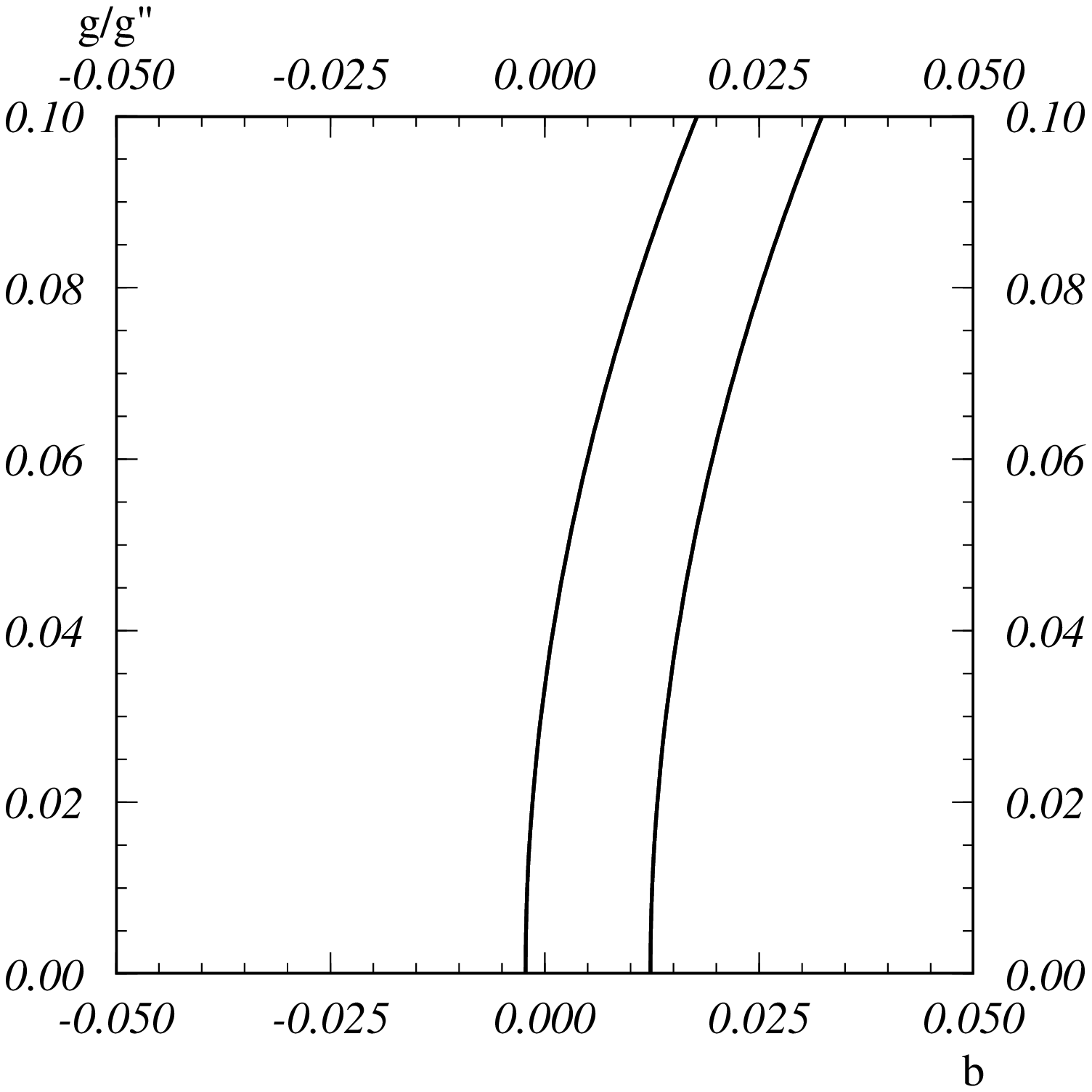}}
\fcaption {$90\%$ C.L. contour in the plane $(b,g/\gs)$ in the limit
of large $M_V$, from LEP data (Glasgow Conference 1994), SLD and low energy
data, with $m_{top}=174$ GeV, $\alpha_s = 0.120$, and $\Lambda=1$ TeV.
The limits on the parameter space are obtained eliminating the heavy vector
field by the use of its classical equations of motion, in the infinite mass
limit. After the elimination of the heavy degrees of freedom the additional
terms to the Standard Model lagrangian allow to read directly the deviations.}
\end{figure}
\smallskip
On the other hand, the process of $W$-pair production by $e^+e^-$ annihilation
would allow for sensitive tests of the strong sector \cite{bessee}, especially
if the $W$ polarizations are reconstructed from their decay distributions.
The importance of this decay channel is due to the strong
coupling between the longitudinal $W$ bosons and the new neutral resonance
$V^0$; furthermore in BESS the Standard Model cancellation among the
$\gamma$-$Z$ exchange diagrams and the neutrino contribution is destroyed.
Therefore the differential cross section grows with the energy.
Explicit calculations show that the leading term in $s$ is however suppressed
by a factor $(g/g'')^4$ and, at the energies considered here, it is the
constant term of the order $(g/g'')^2$ that matters.

Final $W$ polarization reconstruction can be done considering one $W$ decaying
leptonically and the other hadronically and it is relevant to constrain the
model, even if already at the level of unpolarized cross section one gets
important restrictions. Assuming an integrated luminosity of $20 \; fb^{-1}$,
$\sqrt{s}=500$ GeV and $b=0$, it is possible to improve the LEP1 limit on
$g/g''$ over the whole $M_V$ range if polarization is measured, up to $M_V
\approx 1$ TeV for unpolarized $W$.

$W^+ W^-$ pairs can be produced also through a mechanism of fusion of a pair of
ordinary gauge bosons, each being initially emitted from an electron or a
positron. This process allows, for a given centre of mass
energy, to study a wide range of mass spectrum for the $V$ resonance, but it
becomes important only for energies bigger than 2 TeV.

In Fig.~2 the restrictions in the plane $(M_V, g/g")$ are given (with $b=0$)
for three different choices of the collider energy, assuming an integrated
luminosity of $20 fb^{-1}$.

\begin{figure}
\epsfysize=11truecm
\centerline{\epsffile{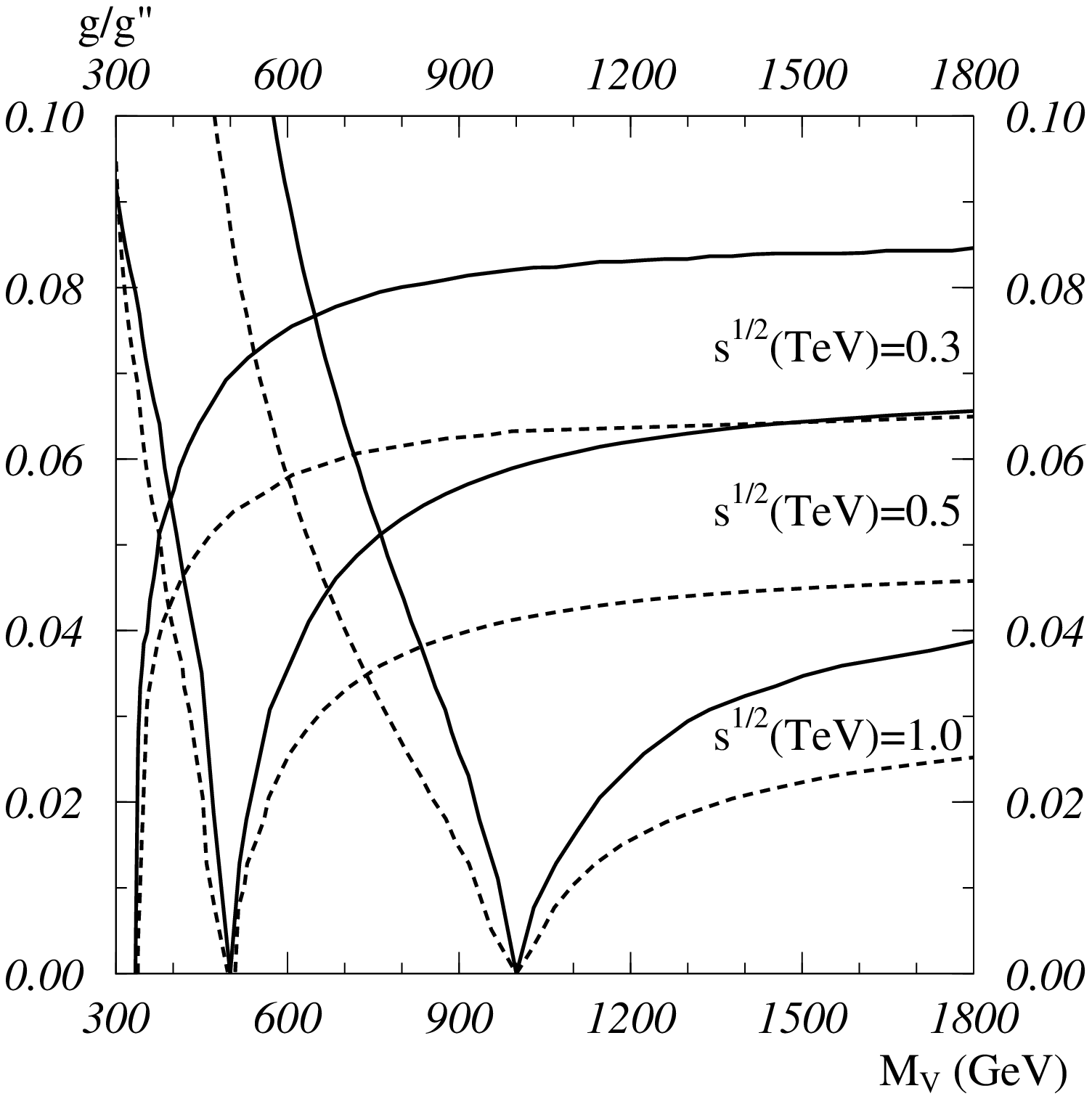}}
\fcaption {$90\%$ C.L. contours in the plane $(M_V,g/\gs)$ for
$\sqrt s=0.3,~0.5,~1~TeV$, $L=20~fb^{-1}$ and $b=0$. The solid line
corresponds to the bound from the unpolarized $WW$ differential cross section,
the dashed line to the bound from all the polarized differential cross
sections $W_{L}W_{L}$, $W_{T}W_{L}$, $W_{T}W_{T}$ combined with the $WW$
left-right asymmetries. The lines give the upper bounds on $g/\gs$.}
\end{figure}
\smallskip

In Fig. 3 the upper bounds on $g/\gs$ for $M_V=1.5$ TeV and $b=0$ are shown
as a function of the center of mass energy of the $e^+ e^-$ collider in the
case no deviation from the SM is found. The relevance of final W polarization
reconstruction over the unpolarized cross section (solid line) is apparent
in the whole energy range considered. A higher luminosity option is shown for
the case $\sqrt{s}=1$ TeV and $L=80~fb^{-1}$ (black dots).

\begin{figure}
\epsfysize=11truecm
\centerline{\epsffile{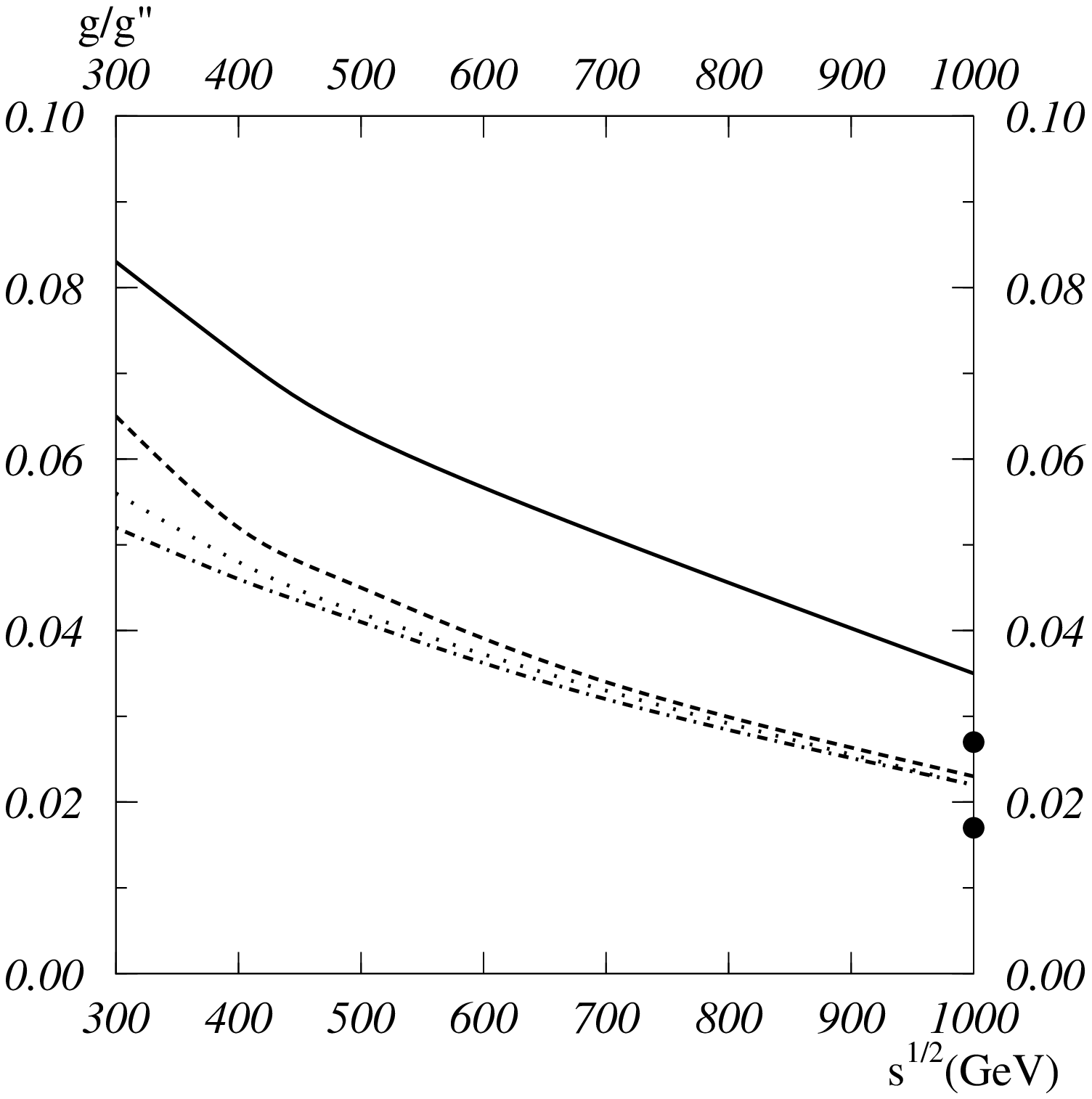}}
\fcaption  {$90\%$ C.L. contours in the plane $(\sqrt{s},g/\gs)$ for
$M_V=1.5~TeV$, $b=0$ and $L=20~fb^{-1}$. The lines correspond to the
unpolarized $WW$ differential cross section (solid line), the $W_{L}W_{L}$
differential cross section (dashed line), and all the differential cross
sections for $W_{L}W_{L}$, $W_{T}W_{L}$, $W_{T}W_{T}$ combined with the $WW$
left-right asymmetries (dotted line) and from all the WW and fermionic
observables with $P_e=0.5$ (dash-dotted line) and represent the upper bounds
on $g/\gs$. The black dots are the bounds for the unpolarized $WW$
differential cross section and from all the WW and fermionic observables at
$\sqrt{s}=1~TeV$ and $L=80~fb^{-1}$.}
\end{figure}
\smallskip
\baselineskip=14pt

In conclusion $e^+ e^-$ colliders could give the possibility to study the
neutral sector of symmetry breaking; $V^0 - Z$ mixing, $V^0 f {\bar f}$ and
$V^0 W^+W^-$ couplings. This is complementarity to
$pp$ colliders (LHC), allowing to explore $V^{\pm}$ resonances through the
decay channel $W^{\pm}Z$. At proton colliders, the channel
$V^0 \to W^+W^-$ is difficult to study due to background problems, and $V^0 \to
l^+ l^-$ has a very low rate.

Linear $e^+e^-$ colliders give also the possibility to study the production
of pairs of charged pseudo-Goldstone bosons, which are present in the extended
version of the model \cite{PGB2}. They can be produced at the $V$ resonance
through the process $e^+ e^- \to V \to P^+ P^-$, where $P^{\pm}$ are the
lightest charged pseudo-Goldstone bosons. The problem of the evaluation of
their masses was investigated in \cite{masse}; the idea is to consider
effective Yukawa couplings between ordinary fermions and pseudos.
The pseudo-Goldstone mass spectrum can be derived from the one-loop effective
potential. The resulting masses are expected to lie in a range
depending on the masses of the heaviest fermions.

The main decay mode of a charged $P$ is $P^+ \to t {\bar b}$, if the
pseudo-Goldstone is heavy enough. It is therefore necessary to analyze
the final state $P^+P^- \to t {\bar b} {\bar t} b$,
and compare it with the background. There are three background sources: $e^+e^-
\to W^+ W^-$, $e^+ e^- \to ZZ$, $e^+ e^- \to t {\bar t}$.
Tagging one $b$ in the final state easily reduces the background $e^+e^- \to
W^+W^-$, while the other two sources are smaller than the signal, at least in
a reasonable range of the model parameter space.

\section{BESS at LHC}

The physics of large hadron colliders has been extensively discussed in a
number of papers (see for example \cite{LHC} and references therein); for what
concerns BESS I shall examine two possible mechanisms to produce $V$
resonances; $q {\bar q}$ annihilation and $WW (WZ, ZZ)$ fusion. In the first
mechanism a quark-antiquark pair annihilates into a $V$, which decays mostly
into a pair of ordinary gauge bosons because the couplings $V^0 W^+_L W^-_L$
and $V^{\pm} W^{\mp}_L Z_L$  are strong (in BESS there is no coupling
$V^0 Z Z$). This process of annihilation always takes place in BESS, even if
$b=0$, due to the mixing between ordinary and new gauge bosons.

The second mechanism goes through fusion of a pair of ordinary gauge bosons,
both of them initially emitted from a quark or antiquark leg, to give a $V$
resonance decaying into a pair $W^{\pm}Z$ or $W^+ W^-$. The cross section is
obtained by a double convolution of the fusion cross section with the
luminosities of the initial $W/Z$'s inside the quarks and the structure
function of the quarks inside the protons. In the $q {\bar q}$ annihilation
only the convolution with the structure functions of the quarks is needed.
The amplitude of the elementary fusion process is strong in BESS: in fact the
scattering of two longitudinally polarized $W/Z$'s proceeds via the exchange of
a $V$ vector boson with large couplings (of the order $g''$ at each vertex).
As I pointed out before, the interesting channel at proton colliders is
$pp \to W^{\pm}Z + X$, because the $W^+W^-$ channel has a strong background
from $pp \to t {\bar t}+X$ and the $ZZ$ one is not resonant in BESS.

So far I have discussed the effects of the triplet of vector
resonances $V$. The real situation could be more complex: axial-vector
resonances might modify in a relevant way the predictions of the minimal model
with only vectors \cite{axial}. As a general feature, virtual effects and
deviations from the Standard Model coming from the vector and axial-vector
sector tend to cancel each other, and the final effects depend on the
relative weight of the two contributions. In some region of the parameter space
of the model there could be complete cancellations and no deviations from the
Standard Model would be observed, at least at energies below the new
resonances. The discovery of a strong electroweak sector only through virtual
effects and precision measurements could therefore be difficult.
The direct discovery of new resonances at the $TeV$ scale would be in such a
case determinant. This is experimentally easy if the width of the resonance
is not too broad (the width increases with the mass $M_V$ and decreses with
$\gs$), while for higher masses (or smaller $\gs$) the discovery potential is
reduced; in particular the shape of the jacobian peak does not differ from
the background, the signal leading only to an excess of events.

\begin{figure}
\epsfysize=11truecm
\centerline{\epsffile{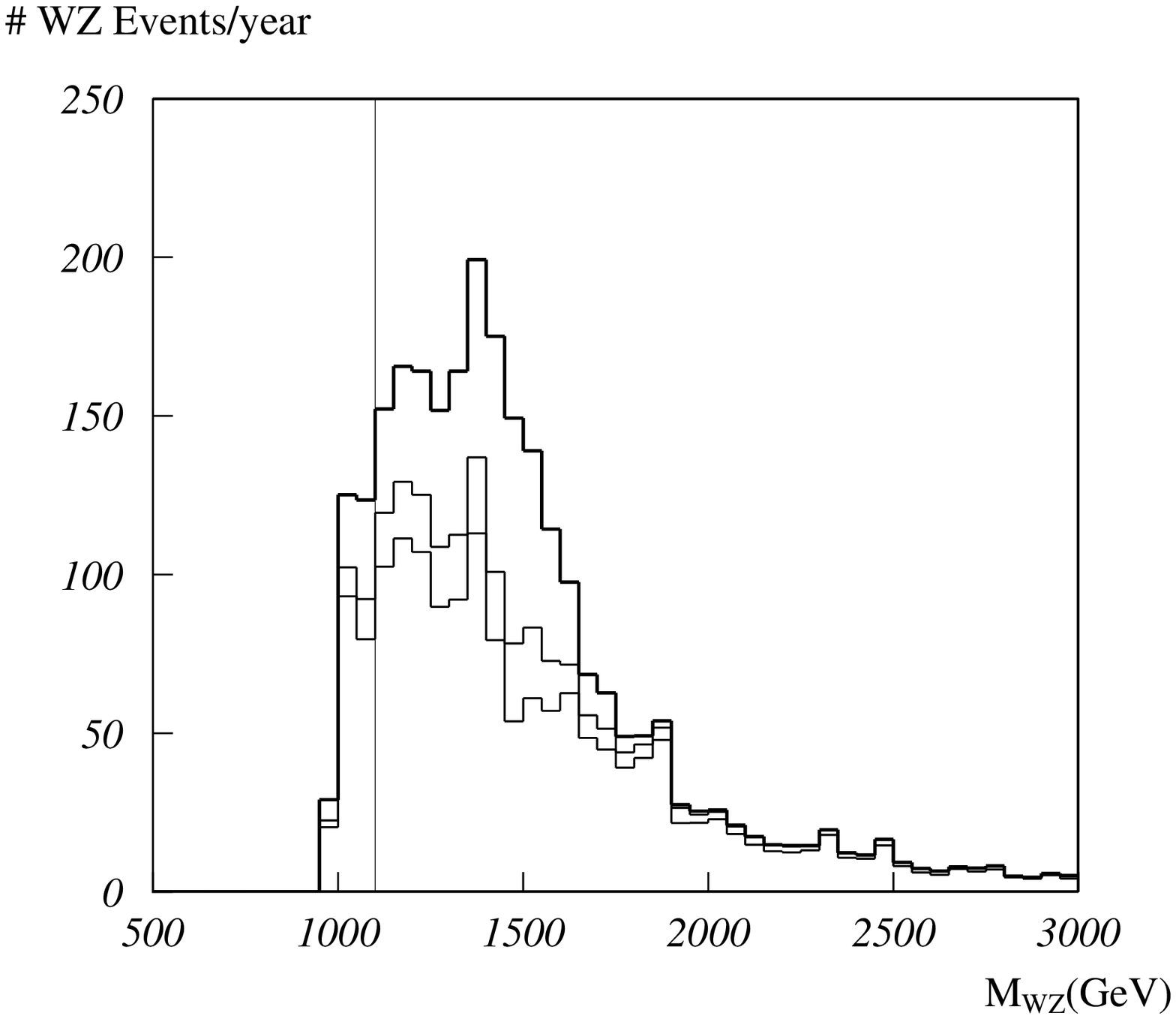}}
\fcaption {Invariant mass distribution of the $W^+Z+W^-Z$ pairs
produced per year at LHC ($\sqrt s =~16$ TeV) for $M_V=1500~GeV$, $g''=13$
and $z=0$ within BESS $SU(8)\otimes SU(8)$, with a luminosity of $100~fb^{-1}$.
The applied cuts are: $|y_{W,Z}|<2.5$, $(p_T)_Z>480~GeV$ and $M_{WZ}>1100~GeV$.
The lower, intermediate and higher histograms refer to the background,
background plus fusion signal and background plus fusion signal plus $q\bar q$
annihilation signal, respectively.}
\end{figure}
\smallskip

In the extended BESS model a richer phenomenology appears. There are $N^2-1$
vector and $N^2-1$ axial-vector new resonances, associated to the local copy of
the global $SU(N)_L \otimes SU(N)_R$. These resonances mix with the ordinary
gauge bosons: in the case $N=8$ the neutral gauge sector involves the mixing of
the fields $W^3, Y,V^3,A^3,V_D$. $V_D$ is a chiral singlet, and its mixing
makes the colorless gauge sector of $SU(8)$-BESS different from the model based
on $SU(2)_L \otimes SU(2)_R$. The $W^{\pm}$, $V^{\pm}$ and $A^{\pm}$ sector is
like in $SU(2)$-BESS. Concerning the colored sector, the $SU(3)_c$ gluons mix
with a color octet of vector resonances $V_8^{\alpha}$

Another new feature is the presence of pseudo-Goldstone bosons.
For quantitative estimates of the
pseudo-Goldstone production cross-sections, I shall employ the
$SU(8)\otimes SU(8)$ extended BESS model. Earlier studies on PGB
phenomenology in technicolor theories \cite{PGB1} \cite{Lubicz} can be found in
the bibliography and references therein. The production is induced by the
processes \cite{PGB2}
\be f^++f^-\rightarrow \gamma,Z, V^3\rightarrow P^+ P^-
\ee
and
\be f_1+f_2\rightarrow W^\pm , V^\pm\rightarrow P^\pm P^0
\ee
where $P^\pm (P^0)$ denote the lightest charged (neutral) PGB's and $f$ denotes
a light fermion.
\begin{figure}
\epsfysize=11truecm
\centerline{\epsffile{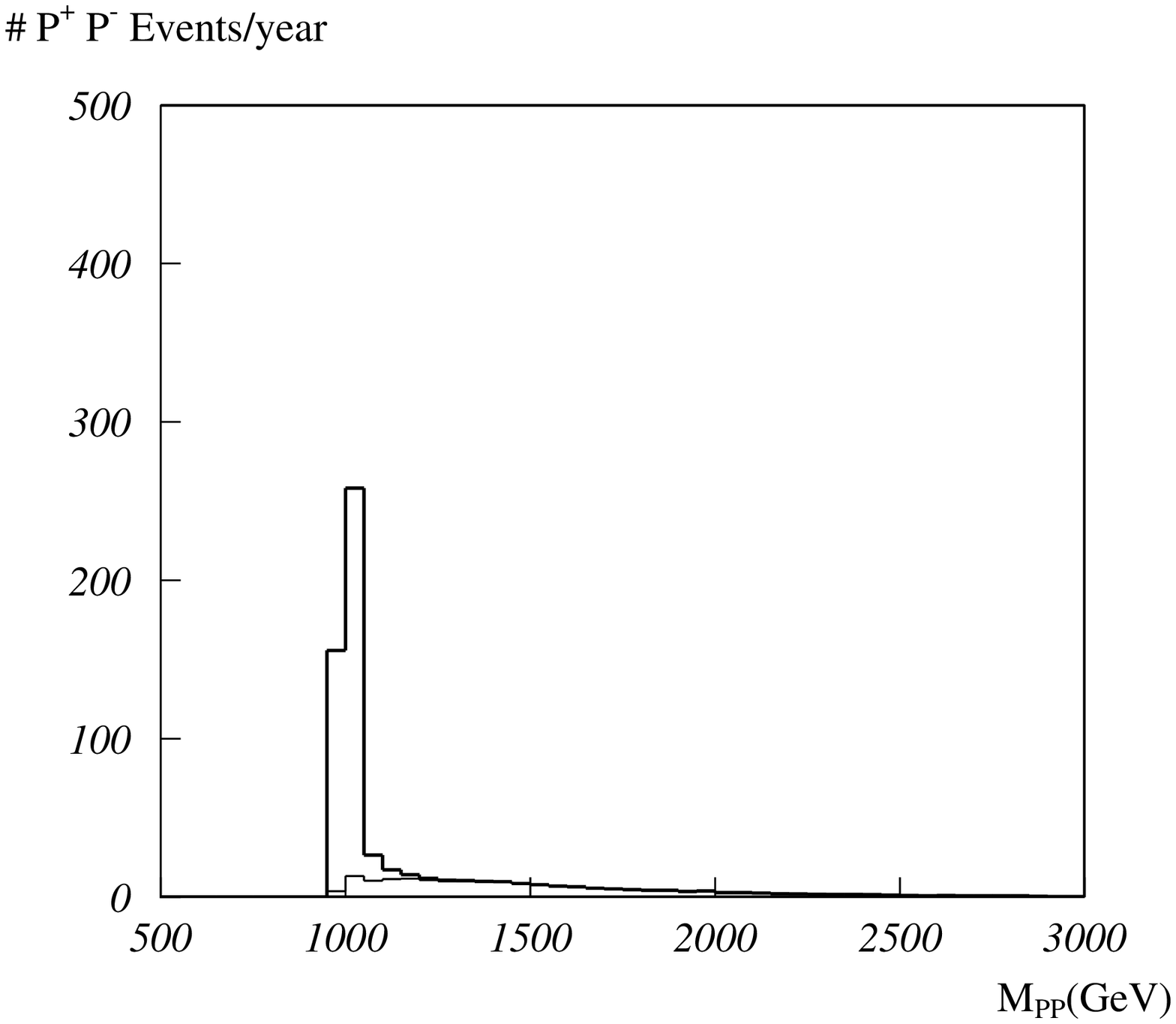}}
\fcaption {Invariant mass distribution of the $P^+P^-$ produced per
year at LHC for $M_P\simeq 400$ GeV, $M_V=1000$ GeV, $g''=13$ and $z=0$,
with a luminosity of $100~fb^{-1}$.
The applied cuts are: $|y_{P}|<2.5$, $(p_T)_{P^0}>300$ GeV.
The lower (higher) histogram refers to the  fusion signal
(fusion signal plus $q\bar q$ annihilation signal).}
\end{figure}
\smallskip
\newpage
\section{Conclusion}

The process of $W$-pair production by $e^+e^-$ annihilation will allow for a
sensitive tests of the strong interacting sector, especially if the $W$
polarizations will be reconstructed from their decay distributions.
The importance of this decay channel is due to the strong coupling between
the longitudinal $W$ bosons and the neutral resonance $V^0$.

The LHC will be able to test the presence of a charged vector resonance
from a strong interacting Higgs sector through its decay into $WZ$
pairs in a significant domain of the extended BESS model parameter space.

Production of pairs of pseudo-Goldstone bosons $P^\pm P^0$ is also important,
but discovery via $tbbb$ or $ttgg$ decays needs a careful evaluation of
backgrounds in the LHC environment.

A more promising possibility is the production of charged
pseudo-Goldstones at the $V$ resonance in $e^+e^-$ collisions in the TeV range.
In fact the largest background, namely $WW$ production, can be easily reduced
to a very low level by requiring the tagging of one $b$ in the final state.
Other backgrounds, such as $ZZ$ and $t\bar t$ production, have smaller
cross-sections as compared with signal cross-section, at least in a range of
the parameter space of the model. For increasing values of the $M_V$ mass and
decreasing values of the $z$ parameter the signal cross-section becomes
smaller than background and deserves a detailed study of background rejection.

\section {Acknowledgments}

I would like to thank R.~Casalbuoni, P.~Chiappetta, S.~De Curtis,
N.~Di~Bartolomeo, D.~Dominici, F.~Feruglio and R.~Gatto for the collaboration
on which this work is based and D.~Dominici also for kindly providing Fig.~1 in
the text. This work is partially supported by the Swiss National Fund.

\end{document}